\documentclass[12pt]{article}
\usepackage{times,amsmath,mathptm,epsfig}
\usepackage{graphicx}
\newcommand\bea{\begin{eqnarray}}
\newcommand\eea{\end{eqnarray}}
\begin{document}
\baselineskip=24pt

\begin{center}
{\large {\bf Structural and Electronic Instabilities in Polyacenes: Density
Matrix Renormalization Group Study of a Long--Range Interacting Model }}\\
\vspace{0.5cm}
C. Raghu$^1$, Y. Anusooya Pati$^2$ and S. Ramasesha$^3$\\
Solid State and Structural Chemistry Unit\\  
Indian Institute of Science, Bangalore 560 012, India\\
\vspace{1cm}
{\bf ABSTRACT\\}
\end{center}
We have carried out Density Matrix Renormalization Group (DMRG) calculations 
on the ground state of long polyacene oligomers within a Pariser--Parr--Pople
(PPP) Hamiltonian. The PPP model includes long--range electron correlations
which are required for physically realistic modeling of conjugated polymers. 
We have obtained the ground state energy as a function of the dimerization 
$\delta$ and various correlation functions and structure factors for $\delta
=0$. From energetics, we find that while the nature of the Peierls' instability 
in polyacene is conditional and strong electron correlations enhance the 
dimerization. The {\it cis} form of the distortion is favoured over the {\it 
trans} form.  However, from the analysis of correlation functions and 
associated structure factors, we find that polyacene is not susceptible
to the formation of a bond order wave (BOW), spin density wave (SDW) or a 
charge density wave (CDW) in the ground state. 

\section {\bf Introduction}

The question whether an infinitely long linear polyene (polyacetylene, PA)
would have equal bond lengths or not has been debated ever since it was known 
that benzene has equal bond lengths while 1,3,5--hexatriene has alternating 
short (double) and long (single) bonds. Addressing this issue, 
Lennard--Jones \cite{lj} and Coulson \cite{coul} predicted a ground state 
with uniform bond lengths. Later work by Labhart \cite{lab}, Ooshika 
\cite{oosh}, Longuet--Higgins and Salem 
\cite{lhs1} established that the ground state would have bond--alternation 
i.e. the infinite polyene would have alternate long and short bonds. 
Experimentally, such a dimerization would qualitatively explain the finite 
optical gap obtained by extrapolation of optical data for linear polyenes 
\cite{kuhn}. Earlier, Peierls established a more general result, 
known as the Peierls' instability or Peierls' distortion \cite{peir} for a 
partially filled one--dimensional band. Peierls' demonstrated that lattice 
vibrations couple to electrons in the band leading to the opening of a gap 
at the Fermi surface, making the material insulating. The H\"uckel model 
solution of Longuet--Higgins and Salem pertain to the specific case of 
Peierls' distortions in a half--filled extended system. Longuet--Higgins 
and Salem showed that for an infinite polyene chain, the total electronic 
energy per carbon atom as a function of the distortion $\delta$ in the chain, 
within a H\"uckel model \cite{huck} is proportional to $\delta^2~{\rm ln}~
|\delta|$ \cite{lhs1}. The elastic strain energy of the system is 
proportional to $\delta^2$, the proportionality constant depending directly 
on the lattice stiffness and inversely on the electron--phonon coupling 
constant. The algebraic forms of the strain and electronic energies 
guarantee that the gain in electronic energy always exceeds the  strain 
energy and the ground state would correspond to nonzero $\delta$. The 
distortion of the polyene chain is termed unconditional as the distorted 
state is always more stable than the undistorted state for any value of 
the lattice stiffness and the electron--lattice coupling constant.
In the last two to three decades, a lot of interest has been generated by 
the possibility of solitonic and polaronic excitations in polyacetylene (PA) 
following the work of Pople and Walmsley \cite{wam}, and later by Rice 
\cite{rice} and Su, Schrieffer and Heeger \cite{ssh}. The latter work 
resulted in the Su--Schrieffer--Heeger (SSH) model for polyacetylene. 

All the work described above was done in the context of non--interacting 
quantum cell models. However, real materials have interacting electrons 
and the importance of including electron--electron interactions has been 
demonstrated while interpreting the ordering of excited states in polyenes
\cite{sr1}. The correct ordering, namely the $2^ {1}A_g$ state lying below 
the optically allowed $1^{1}B_u$ state in long polyenes can be obtained 
only in a correlated electronic model. The effect on dimerization in the 
ground state of polyacetylene in a correlated model was unclear till the 
real--space, valence--bond (VB) study by Mazumdar and Dixit \cite{sumit1} 
for Hubbard model and Soos and Ramasesha for PPP model \cite{sr0}. While 
mean--field and Hartree--Fock approaches predicted a decrease in dimerization, 
Mazumdar and Dixit found a clear enhancement, over a fairly large range of 
correlation strength with a maximum dimerization around $U \sim 4t$. In 
their study they employed a model that included electron correlation at a 
minimal level, namely the Peierls'--Hubbard model. This result was 
subsequently verified by quantum Monte Carlo study by Hirsch \cite{hirs}, 
variational calculation by Baeriswyl and Maki \cite{maki} and numerical 
renormalization group study by Hayden and Mele \cite{mele}. Soos and 
Ramasesha \cite{sr0} found a similar result in case of PPP model which includes 
extended range electron correlation. Thus it is now widely believed that 
electron correlations enhance dimerization in the Peierls--Hubbard model.

The Peierls' instability is mainly a one--dimensional phenomena since only 
in one--dimension does the Peierls' gap open along the full Fermi surface.
In higher dimensions, Peierls' instability is not realized as easily as in 
one--dimension \cite{peir}. The effect of dimensionality on the extent of 
instability has been studied within the frame work of energy band theory, 
where it is held that the strength of the instability depends on the extent
of nesting of the Fermi surface. 
The effect of dimensionality on the instability, in going from one--dimension 
to quasi one--dimension can be explored by studying coupled chain (ladders) 
systems. The realm of quasi one--dimension is interesting as it provides for 
interplay of strong quantum fluctuations as seen in one--dimension on the one 
hand and possibility of charge or spin to order as seen in two--dimensions
on the other. Historically, ladders have been investigated within the 
Hubbard model mainly to explore instabilities like pairing instability, 
charge and spin gaps \cite{srw}, or charge ordering phenomenon \cite{noack}. 
The experimental realization of these systems are the two--legged 
${\rm SrCu_2O_3}$ and the three--legged ${\rm Sr_2Cu_3O_5}$ ladder compounds 
\cite{lads} wherein the spin exchange or electron transfers are confined to 
a pair of parallel chains. 

In the realm of conjugated polymers, polyacenes (Fig. 1a) can be viewed as 
ladders with missing alternate rungs. These systems can be fairly easily 
realized experimentally. Polyacene molecules with up to seven rings have already 
been synthesized in the laboratory. Recently, polyacenes received a lot of 
attention when tetracene was used as the lasing material to make the first 
electrically pumped organic laser \cite{bat1}. It was also demonstrated that 
one could make very high quality field--effect transistors (FETs) based on 
pentacene.  These devices display exotic behaviour like fractional quantum hall 
effect, superconductivity and electrical switching in the FET configuration 
\cite{bat2}. Quantum chemists have been interested in polyacenes for a 
very long time. Early studies on Peierls' instability in polyacenes were 
carried out by Salem and Longuet--Higgins within the H\"uckel model 
\cite{lhs2}. They observed that unlike in the case of PA, instability in 
polyacene is conditional; it depends on the lattice stiffness for a 
given electron--phonon coupling. In PA, the distorted state is more stable 
than the undistorted state for any value of the force constant and 
electron--phonon coupling. Salem and Longuet--Higgins considered the {\it cis} 
distorted form (see Figs. 1b and 1c) while Boon \cite{boon} argued that 
the {\it trans} form was the more stable form of distortion. Subsequent 
studies by Whangbo, Woodward and Hoffmann \cite{hoff}, and Tanaka {\it 
et al.} \cite{tana} added to this claim. Studies including electron 
correlations were initially confined to the mean--field picture. Kivelson 
and Chapman \cite{kiv} studied bond alternation, magnetic ordering and 
possible superconductivity in polyacenes. In addition, there have been 
various other studies on the origin of the band gap, metal--insulator 
transition and spin--Peierls' distortions \cite{poly1}. The antiferromagnetic
spin--1/2 Heisenberg system with nearest neighbour exchange corresponding to 
polyacene geometry has been studied by Garcia--Bach, Valenti and Klein to 
explore the possibility of spin--Peierls' instability in the system 
\cite{garcia}.

Studies of Peierls' instability using quantum cell models with explicit
electron correlations in the context of polyacenes have been very few and
these investigations have employed only the Hubbard model.
A modified Gutzwiller variational study by O'conner and Watts--Tobins 
\cite{con} reaffirmed the conditional nature of the instability. The 
instability investigated by employing the Projector Quantum Monte Carlo 
technique (PQMC) for finite oligomers of polyacenes concluded the 
same \cite{sr2}. This study also found that electron correlations 
enhance the susceptibility to distortion. Besides, for large interaction 
strength, the undistorted polyacene was found to show a tendency for
formation of a SDW ground state. 

Polyacenes, being semiconducting, the electron--electron interactions are 
long ranged. The QMC technique is difficult to implement for quantum cell 
models with long range interactions. The Hubbard--Stratanovich transformation 
of each interaction term leads to a bosonic variable in QMC. Thus, even for 
small quantum cell systems, the number of bosonic variables become 
prohibitively large. Besides, the QMC method is also restricted to intermediate
correlation strengths. In our present study of instability in polyacene, we 
have used a Pariser--Parr--Pople (PPP) model and have employed the Density 
Matrix Renormalization Group (DMRG) method \cite{white} to solve the model. The 
PPP model includes long--range coulomb interactions and appears unsuitable for 
a DMRG study as it apparently spoils the quasi--one--dimensional topology
of the system. However, our earlier studies on PA has shown that the DMRG 
method is quite accurate for the PPP model \cite{drsdn} and it appears that 
the DMRG method retains its accuracy if the long--range interaction part is 
diagonal, as is indeed the case in the PPP model. In this paper we first 
review the results obtained from a non--interacting model, then present our 
results and discussion on polyacene and finally a summary of the results. 

\section {\bf The non--interacting picture}

The analysis by Salem and Longuet--Higgins \cite{lhs2} was based on a H\"uckel 
model study of polyacene. They considered an infinitely long chain of polyacene 
as made of two infinite polyenes joined together by cross links (rungs), as 
shown in Fig. 1a. The molecular orbitals (MOs) of each unit cell of  polyacene 
may be classified as symmetric or antisymmetric according to their symmetry 
with respect to reflection about the plane bisecting the rung bonds, indicated 
in the figure by a dashed line. There are in all four bands, arising from the 
four $\pi$--MOs in each unit cell, two of each symmetry type as shown in Fig. 
2. In the undistorted form of polyacene, there is no band gap between the
occupied and empty bands, while in both the {\it cis} and {\it trans} distorted
forms (Figs. 1b and 1c) there is gap between the occupied and empty bands. 
In the undistorted case, since the lower two bands are completely filled
and the upper two bands are empty, we expect the system to be insulating. 
However, the absence of a band gap between the filled lower band and the empty 
higher band, due to accidental degeneracy, makes uniform polyacene a rather 
unconventional metal. If a symmetrical distortion potential is imposed by
way of a dimerization one might expect this degeneracy to be lifted. However,
the matrix element of the perturbation between the symmetrical and 
antisymmetrical states at the band edge vanish by symmetry. Thus the case of 
polyacene, even in the noninteracting limit, is different from PA and we cannot
conclude if we will observe a dimerization of the chain. Salem and 
Longuet--Higgins \cite{lhs2} conjectured that the polyacene would show only a
conditional instability.

Kivelson and Chapman \cite{kiv}, based on their study of polyacene by a 
nearest--neighbour tight binding model, argue that since the energy of the
states near the Fermi energy is a quadratic function of the wave vector
$k$ (rather than linear as in PA) the density of states diverge near the Fermi
energy, enhancing the possibility of instabilities in polyacene. Their study
arrives at the conclusion that there is no structural instability of the two
kinds depicted in Figs. 1a and 1b. They speculate on the possibility of a 
superconducting or a magnetically ordered ground state. They however ignore
electron--electron interactions by considering them as "probably weak".

\section {\bf The PPP model and Computational Scheme}

The PPP model has been widely studied in the context of conjugated polymers
and molecules \cite{sumit2}. The PPP model Hamiltonian is given by,

\bea
{\hat H}_{\rm PPP}~=~{\hat H}_0+~{\hat H}_{{\rm int.}}~~~~~~~~~~~~~~~~~~~~~~~~~~~~~~~~~~~~~~~~~~~~~~~~~~\\  \nonumber
{\hat H}_0=\sum_{l=1}^{2} \sum_{i,\sigma}t_0[1+\eta_{l}(-1)^{i}
\delta](\hat a_{i,l\sigma} ^ \dagger \hat a^{}_{i+1,l\sigma}+H.c.)
+ \sum_{i,\sigma}t_0(\hat a_{2i-1,1\sigma} ^ \dagger \hat a^{}_{2i-1,2\sigma}
+H.c.),\\
\nonumber
 {\hat H}_{{\rm int.}}~=~\sum_{i}~\sum_{l}~\frac{U_{il}}{2}{\hat n}_{i,l} 
({\hat n}_{i,l}
 -1)~+~\sum_{i,j}~\sum_{l,m} ~V_{il,jm}~ ({\hat n}_{i,l} - 1) 
({\hat n}_{j,m} - 1)~. 
\eea
where $l,m$ are the chain index, and $i,j$ refer to sites on a chain,~ $\hat 
a_{i,l \sigma}^\dagger~(\hat a^{}_{i,l \sigma})$ creates (annihilates) an 
electron of spin $\sigma$ at site $i$ on chain $l$, ~$t_0$ is the transfer 
integral, which alternates between $t_0(1~+~\delta)$ and ~$t_0(1~-~\delta)$ for 
adjacent bonds on the same chain,~ $\delta$ being the dimensionless 
dimerization parameter, $\eta_{1}=\eta_{2}=1$ if the dimerization of the 
top ($l=1$) chain is in phase with the dimerization of the bottom chain 
($l=2$) ({\it cis} configuration) and $\eta_{1}=1, \eta_{2}=-1$ if 
the dimerization of the top chain is out of phase with the dimerization of 
the bottom chain ({\it trans} configuration). $U_{il}$ is the on--site Hubbard 
interaction and $V_{il,jm}$ is the intersite Coulomb interaction, interpolated
between $U_{i,l}$ and $e^2/r,~~ r \longrightarrow \infty$, using
the Ohno interpolation scheme \cite{ohno}, 
\bea
V_{il,jm}=14.397\left [{\frac  {28.794} {(U_{il}+U_{jm})^2}}+r_{il,jm}^2
\right]^{-{\frac {1} {2}}} ~,
\eea
(the distances are in \AA ~and the energies in eV) widely used for conjugated
polymers. In all our computations,
we have used the standard PPP parameters for carbon, which is $U=11.26 $~eV,
$t_0=2.4 $~eV for bond length $r=1.397$ \AA $~~$which corresponds to $\delta = 0$.
In computing distances we have assumed that all bond angles are $120^\circ$
and that bond lengths scale as $r(1-\delta)$ for a bond with transfer integral 
$t=t_0(1+\delta)$.

The DMRG method, invented by White \cite{white} is the most accurate numerical 
method yet for calculating the ground and low--lying states of interacting 
quasi--one--dimensional systems. In our application of this technique to 
polyacenes within the PPP model, there are two crucial differences from all 
the earlier implementations of the technique for low--dimensional systems. 
These are (i) the interaction part, although diagonal in the real space 
representation, is truly long--ranged and (ii) the topology of the system being 
constructed does not have structures corresponding to oligomers of polyacene 
at every stage of the DMRG iteration. Incorporating long--ranged interaction 
implies that we need to renormalize and store the matrices corresponding to 
the number operators of all the sites at each iteration. While this is quite 
straightforward, it takes up large storage and for efficiency of computations 
we have stored them in sparse form. Typically only about 5\% of the matrix 
elements of the site number operators are nonzero. 

The nonlinear topology of polyacenes implies that we should find an efficient
and accurate way of building the oligomers. We should avoid using cyclic
boundary conditions or long range transfer operators in building the final
system. In Fig. 3 we show schematically the way the polyacene oligomers are
constructed in the DMRG scheme. We begin with a ring of four sites and 
systematically add new sites in the middle of the system building up the 
molecule in such a way that we avoid (i) long--range transfers between new 
and old sites and (ii) transfers between old sites that are far apart. We 
have extensively checked this procedure by carrying out calculations on large 
oligomers in the noninteracting limit and comparing them with exact numerical
results. 

We have performed both infinite and finite DMRG calculations on polyacene 
chains with up to $24$ monomer units. We have done DMRG calculation with
$128$ density matrix eigenvectors after making sure that the energies
converge for this truncation, by calculating for a few oligomers with 
larger cut--offs. Besides, the ground state energy for napthalene and 
anthracene obtained from DMRG with a cut--off of 128 compare very well 
with model exact calculations in a Valence Bond (VB) basis \cite{sr4,sr5}. 
Symmetries 
like conservation of $z$--component of total spin $S_z$ and total number 
of electron $N_{tot}$ have been exploited in our implementation. Since 
the left block--right block reflection symmetry commonly seen in 
DMRG calculations is broken due to the bond alternation in each chain, 
we have to store relevant operators for every site on both the left 
and right blocks. This doubles the storage but can be easily handled 
as all the site operators are highly sparse. Properties like expectation 
value of observables or correlation functions can be evaluated with 
great accuracy after a few iterations of finite system DMRG algorithm. 

\section {\bf Results and discussion}

Our studies are divided into two parts: (1) study of the dependence of the 
energy of the system on $\delta$, and (2) study of susceptibility of the 
ground state to instabilities for $\delta~=~0$ as seen from appropriate
structure factors.  We approach the question of Peierls' distortion first 
on the basis of energetics.  We have calculated the total energy of the 
system as a function of the system size $N$ and dimerization $\delta$, 
up to a maximum system size of $24$ monomer units for $\delta$ values 
ranging from 0.01 to 0.1 by employing the infinite system DMRG algorithm. 
In Fig. 4 we show the dependence of the energy per unit cell on the inverse 
system size, $1/N$. We note that this dependence is linear in $1/N$. In the 
case of the {\it trans} polyacene, oligomers with even and odd number of 
unit cells fall on two different straight lines for large $\delta$. The 
smooth behaviour of the energy per unit cell {\it versus} $1/N$ gives 
confidence in extrapolation of the enegries to the thermodynamic limit. 

In order to study the nature of the structural instability, if any, in the 
polyacenes, we study the stabilization energy for the distorted structure, 
$\Delta E_{A} (N,\delta)~=~$ $E(N, 0)~-~ E_{A}(N,\delta)$ where $A~=~C~ 
{\rm or}~ T$ corresponding to {\it cis} or {\it trans} form of polyacene, 
for various $\delta$ and number of rings, $N$ in the polyacene. By definition, 
positive values of $\Delta E_{A}$ would indicate that the distorted structure 
is stabilized. We obtain the stabilization energy in the thermodynamic 
limit, $\Delta E_{A} (\infty,\delta)$, for each value of $\delta $ from  
$\delta ~=~0.01$ to $0.1$ by extrapolation of $\Delta E_{A} (N,\delta)/N$ 
{\it versus} $1/N$. In Fig. 5 we present the variation of $\Delta E_{A}
(\infty,\delta)$ with $\delta$. Also shown for comparison are the H\"uckel 
results for this system. We note that the stabilization of the dimerized 
state in the interacting system is larger than the stabilization of the 
same in the noninteracting model for both {\it cis} and {\it trans} 
distortions. This shows that dimerization is favoured in the interacting 
models more than in the noninteracting models, just as in the case of polyenes. From the nature of the curve in Fig. 5, the stabilization energy appears 
to be quadratic in $\delta$, for both {\it cis} and {\it trans} form of 
polyacene. The plot of $\Delta E_{A}(\infty,\delta)$ {\it vs} $\delta^2$
shown in the inset, in fact shows that the stabilization is quadratic in 
$\delta$ to a very good approximation. The quadratic form of the electronic
stabilization of the dimerization implies that the instability in polyacenes 
is indeed conditional for both the {\it cis} and the {\it trans} forms. This 
is in contrast to the unconditional dimerization seen in PA. Furthermore, it 
is also seen that for the same distortion $\delta$, the {\it cis} form is more 
susceptible than the {\it trans} form to dimerization. Earlier work \cite{sr2} 
within a Hubbard model also found that the {\it cis} form is stabilized more
than the {\it trans} form in the presence of electron--electron interactions. 
The long range nature of interactions considered here does not seem to affect 
this result. Indeed, even the H\"uckel model also shows that the {\it cis} 
form is stabilized more than the {\it trans} form.

To get a physical picture of the ground state, we have studied many different 
types of correlation functions. In the context of a structural instability
like the formation of a bond--order wave (BOW), the bond--bond correlation 
function of the undistorted system gives information about the type of BOW 
instability that may occur in the system. The Fourier transform of the 
bond--bond correlation function, the bond structure factor, gives the amplitude 
for various BOW instabilities that may be present in the ground state. 
Similarly, the structure factors associated with charge--charge and spin--spin 
correlation functions provide information on the susceptibility of the system 
towards CDW and SDW instabilities. We have calculated bond--bond, spin--spin 
and charge--charge correlation functions in the ground state for twenty four 
ring polyacene system which is the largest system size that we have been able 
to study. A unit cell of polyacene has five bonds as shown in Fig. 6a; of 
these the bonds 1, 2, 4 and 5 are important to characterize a distortion as 
either {\it cis} or {\it trans}. It is possible to construct five different 
types of bond-bond correlation functions with respect to a reference bond. 
From our view point, the important correlation functions involve bond 1 with 
bonds 1 and 2 in each unit cell as well as bond 1 with bonds 4 and 5 in each 
unit cell. The first type of correlation function is useful in understanding 
if the polyacenes indeed distort while the second type are useful in 
identifying the type of distortion if it is present.

The bond--bond correlation function is defined as
\bea
<\hat b_{i,l}~\hat b_{j,l^\prime}>~=~<~(\sum_{\sigma}\hat a_{i,l~\sigma}^
{\dagger}~\hat a_{i+1,l~\sigma}+H.c.)
(~\sum_\tau\hat a_{j,l^{\prime}~\tau}^{\dagger}~\hat a_{j+1,l^{\prime}~\tau}
+H.c.)~>~~,
\eea
where $l,l^{\prime}$ denote the chain. In the DMRG method, the expectation
value of the product of two operators are very accurate, if they belong to
different blocks. Thus, the bond-bond correlations are calculated between the 
bond on one half with all the bonds on the other half. The bond operator is 
itself a product of two operators in the same block. We have computed the 
matrix of this product explicitly when it is first encountered and have 
renormalized the product matrix thereafter. The correlation functions are 
calculated for open polyacene chain. However, the resulting correlation 
function cannot be Fourier transformed since the system is not strictly 
periodic. To overcome this difficulty, we have assumed that the correlations 
involving the interior rings are identical with those computed for a periodic 
system (Fig. 6b). This is reasonable, if we neglect sites belonging to the 
two rings at each of the ends. Then the correlations corresponding to the 
properties of the interior twenty rings are taken to be the same as those in 
a twenty ring system with periodic boundary conditions. This allows obtaining 
structure factors from the corresponding correlation functions. To enhance the 
accuracy of our calculations, we have carried out {\it finite} DMRG 
calculations for the 24 ring polyacene system.  We have kept the DMRG 
cut--off at 150 density matrix eigenvectors and have used four finite DMRG 
sweeps. 

We show the plot of bond-bond correlations for the bond in the bottom chain
(shown in bold in Fig. 6b) with all the chain bonds on the top and bottom 
left half of the system in Figs. 7a and 7b. We note that except at the ends 
of the chain, the correlations are almost identical. The bond-bond correlation 
shows a slight short range oscillation which dies off rapidly. Eliminating the 
end bonds and imposing periodicity as discussed earlier, we have obtained the 
structure factor corresponding to the bond-bond correlations. This is shown in 
Fig. 8. We note that the structure factor has no peaks anywhere away from 
$q=0$. This clearly implies that the system is not susceptible to any bond 
order wave. Since a single chain remains uniform, the question of {\it cis} or 
{\it trans} type of distortion does not arise. We have also confirmed that the 
bond-bond correlation for the rung bonds shows a similar behaviour, implying 
that the rung bonds are uniform.

The spin--spin correlation functions, $<s_{i,l}^zs_{j,l^\prime}^z>$ and 
charge--charge correlation functions, $<n_{i,l}n_{j,l^\prime}>$ have also 
been computed to see if the system has a tendency for formation of
either a SDW or CDW. According to Fig. 9a, there are four sites in a unit 
cell of undistorted polyacene. We have calculated these correlations between
the new right site (Fig. 9b) and all the sites in the left block. 
The real space spin--spin correlations, displayed in Figs. 10a and 10b show 
short range antiferromagnetic fluctuations expected from a nondegenerate 
correlated model \cite{sr3}. The spin--spin correlation decays exponentially 
which is consistent with a spin gap in the system. The charge--charge 
correlations, shown in Figs. 11a and 11b show very slight short range 
oscillation, extending over a couple of sites. They also decay rapidly. The 
structure factors for spin--spin and charge--charge correlations are shown in 
Figs. 12 and 13. They clearly show that the system favours uniform charge 
distribution and no spin density oscillations. This rules out any possibility
of CDW or SDW ground state in polyacenes. 

\section {\bf Summary}

We have studied polyacenes up to $24$ rings with long range coulomb 
interactions 
within a PPP model, by the DMRG method. From energetics, we conclude that the 
structural instability in polyacene is only conditional, unlike that in the 
case of PA. We find that the {\it cis} form of distorted polyacene is more 
stable than the {\it trans} form. This is contrary to earlier predictions based 
on non--interacting models where the {\it trans} form was predicted to be
more favourable. We have used the finite DMRG algorithm to calculate 
correlation functions like bond--bond, spin--spin and charge--charge in the 
undistorted ground state. Analysis of these correlation functions and their 
associated structure factors leads to the conclusion that polyacene ground 
state does not have the tendency to show BOW, SDW or CDW instability. 

\noindent{\bf Acknowledgements:}
We acknowledge financial support of CSIR (Project No. 1595/99/EMR-II) and 
BRNS, Dept. of Atomic Energy, (Project No. 99/37/37/BRNS), India. \\
\noindent
$^1$raghu@sscu.iisc.ernet.in ;
$^2$anusooya@sscu.iisc.ernet.in ; 
$^3$ramasesh@sscu.iisc.ernet.in

\begin{figure}
\begin{center}
\epsfig{figure=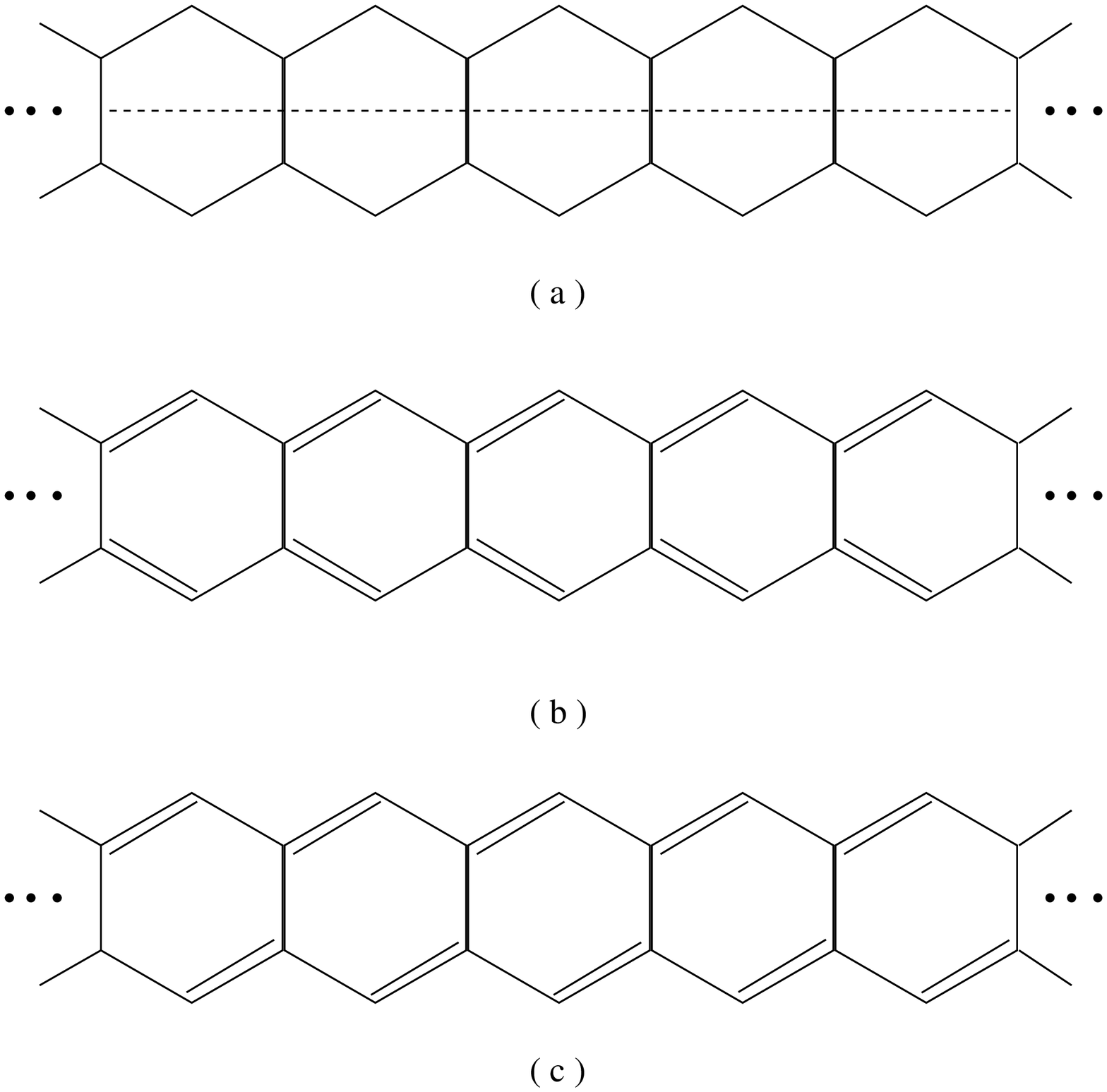,height=16cm,width=15cm}
\caption[] {Structure of polyacene. (a) Undistorted or uniform, 
(b) Cis distorted form, (c) Trans distorted form.}
\label {fone}
\end{center} 
\end{figure}

\begin{figure}
\centerline{\includegraphics[height=12cm]{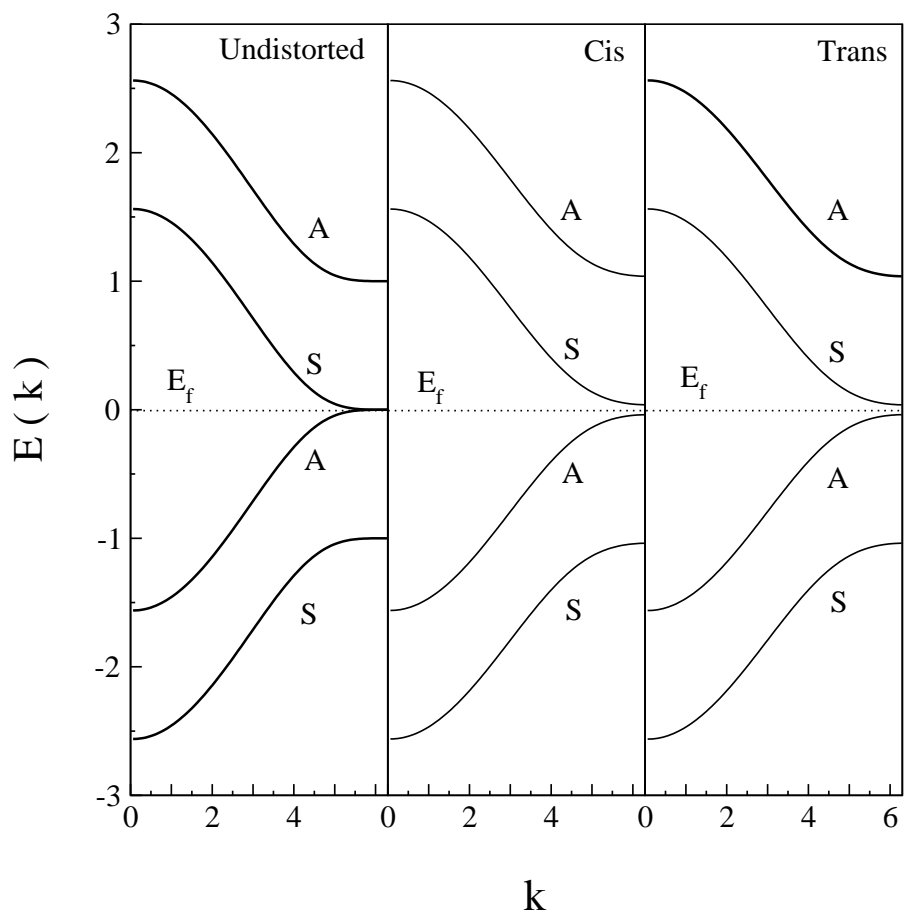}}
\begin{center}
\caption{Dispersion of the $\pi$--bands in polyacene for the three forms.
 The {\it cis} and {\it trans} forms correspond to dimerization of 
 $\delta~=~0.2$.}
\end{center} 
\end{figure}

\clearpage

\begin{figure}
\centerline{\includegraphics[height=16cm]{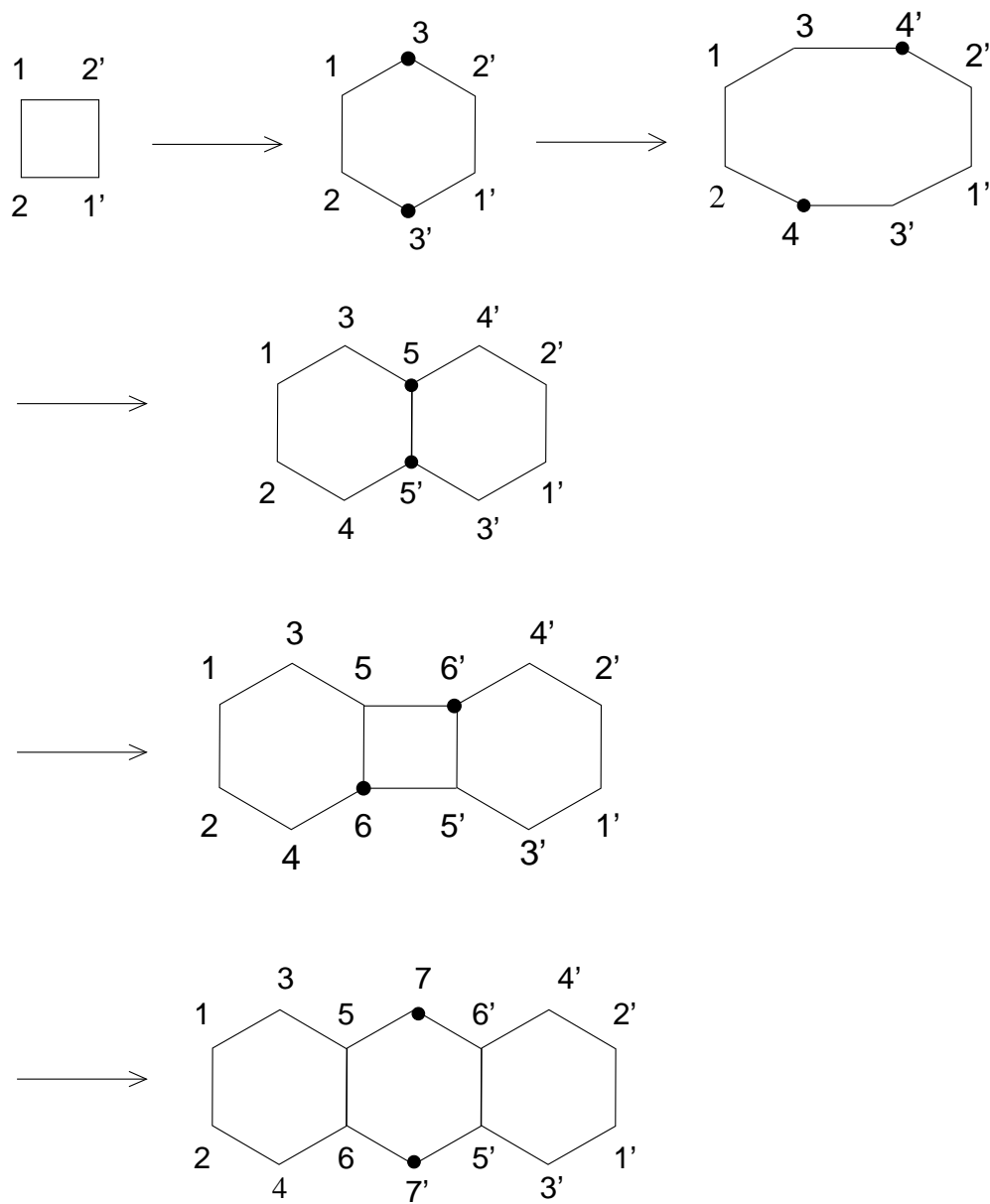}}
\begin{center}
\caption{Inside-out scheme for building polyacene oligomers, adding two sites
 at a time in the DMRG procedure, starting from a four site system. The primed 
 sites correspond to the right block and the unprimed sites correspond to the
 left block.}
\end{center} 
\end{figure}

\clearpage

\begin{figure}
\centerline{\includegraphics[height=15cm]{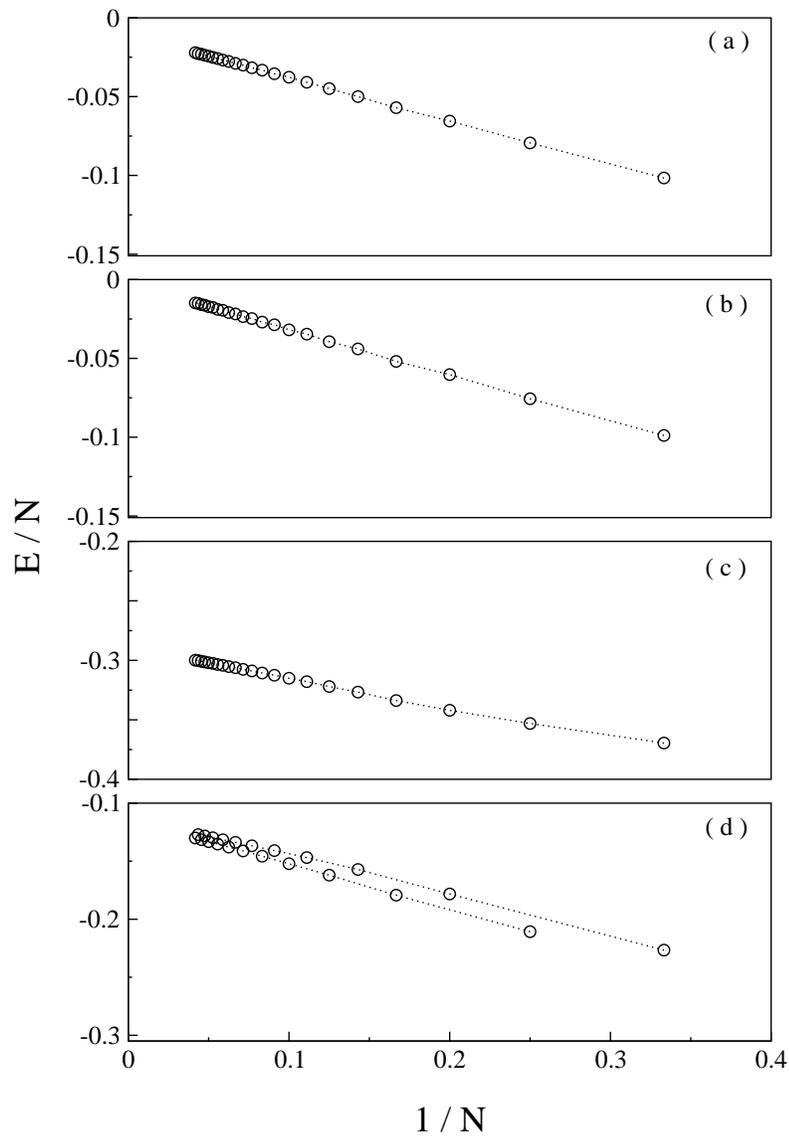}}
\begin{center}
\caption{Convergence of energy per unit cell for polyacene oligomers, in the
 DMRG calculation, for (a) cis, $\delta~=~0.01$, (b) trans, $\delta~=~0.01$, 
(c) cis, $\delta~=~0.1$, (d) trans, $\delta~=~0.1$.}
\end{center} 
\end{figure}

\clearpage

\begin{figure}
\centerline{\includegraphics[height=12cm]{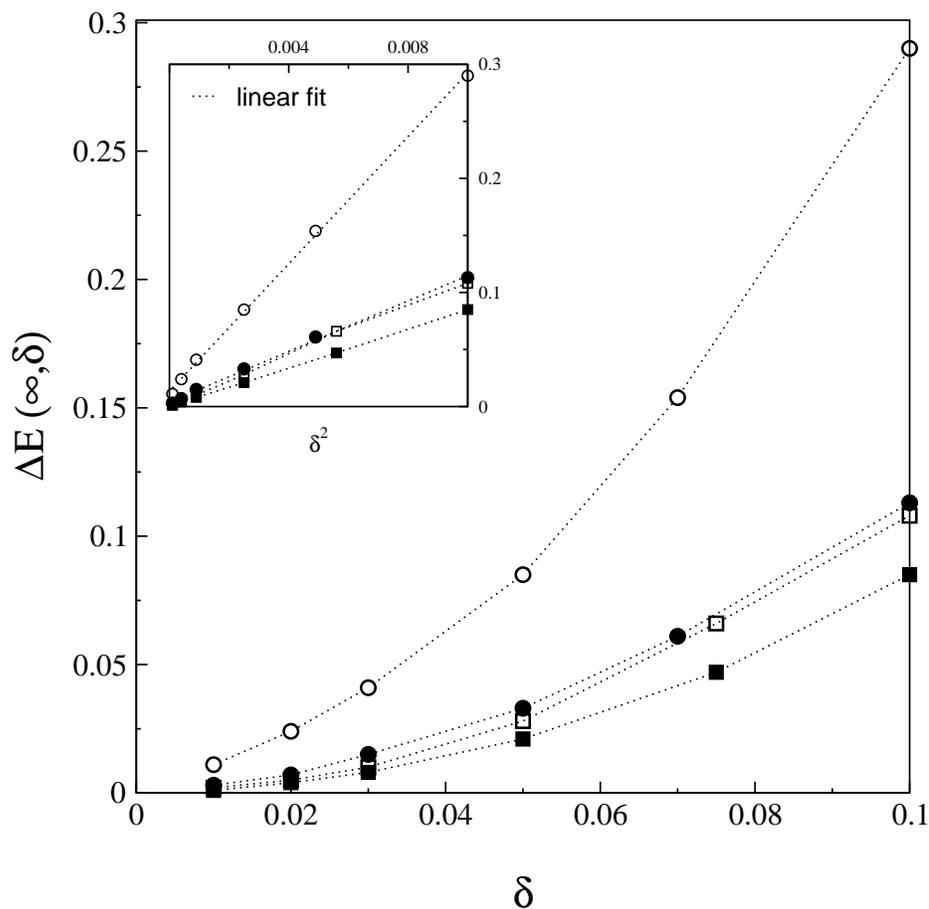}}
\begin{center}
\caption{Stabilization energy for dimerization, as a function of $\delta$, the
 distortion. Circles and squares are for PPP and H\"uckel calculations and 
 open and closed symbols stand for {\it cis} and {\it trans} forms of 
 polyacene respectively. The inset figure shows the linear dependence of the 
 stabilization energy when against $\delta^2$.}
\end{center} 
\end{figure}

\clearpage

\begin{figure}
\centerline{\includegraphics[height=4cm]{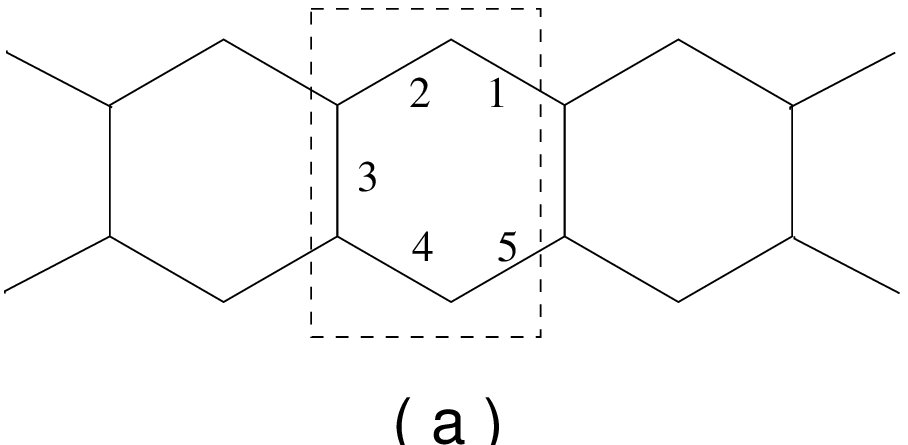}}
\vspace{2.0cm}
\centerline{\includegraphics[height=3.5cm,width=15cm]{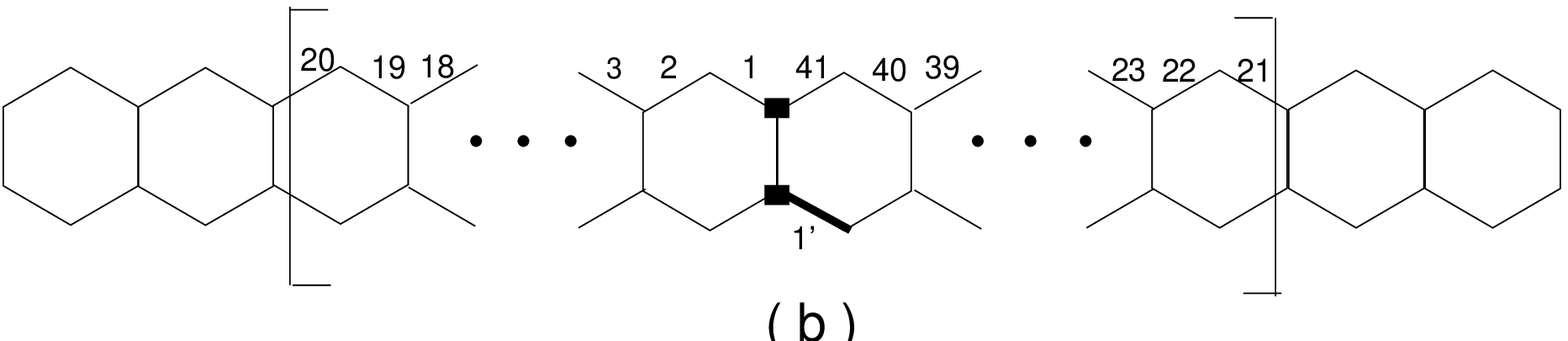}}
\vspace{0.5cm}
\begin{center}
\caption{(a) Unitcell of polyacene showing the different bonds, (b) numbering 
 of bonds in polyacene chain. The new right bond indicated by a thick line is 
 the reference bond for bond-bond correlation. The new sites are indicated by  
 filled squares. The square brackets indicate the part of the system over 
 which periodic boundary conditions are applied to calculate bond structure 
 factors.}
\end{center}
\end{figure}

\clearpage

\begin{figure}
\centerline{\includegraphics[height=14cm]{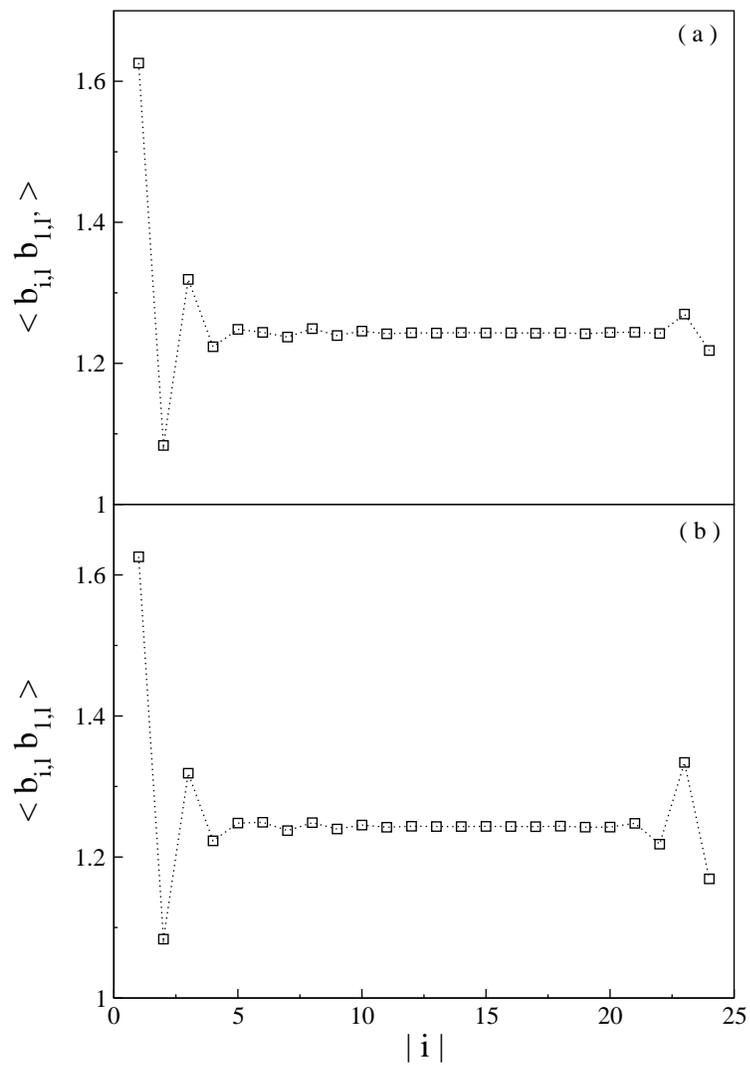}}
\begin{center}
\caption{Bond--bond correlation as a function of the seperation between bonds,
 between the bond shown in bold in Fig. 6b and the bonds in (a) the upper
 chain and (b) the lower chain of the left block.}
\end{center}
\end{figure}

\clearpage

\begin{figure}
\centerline{\includegraphics[height=12cm]{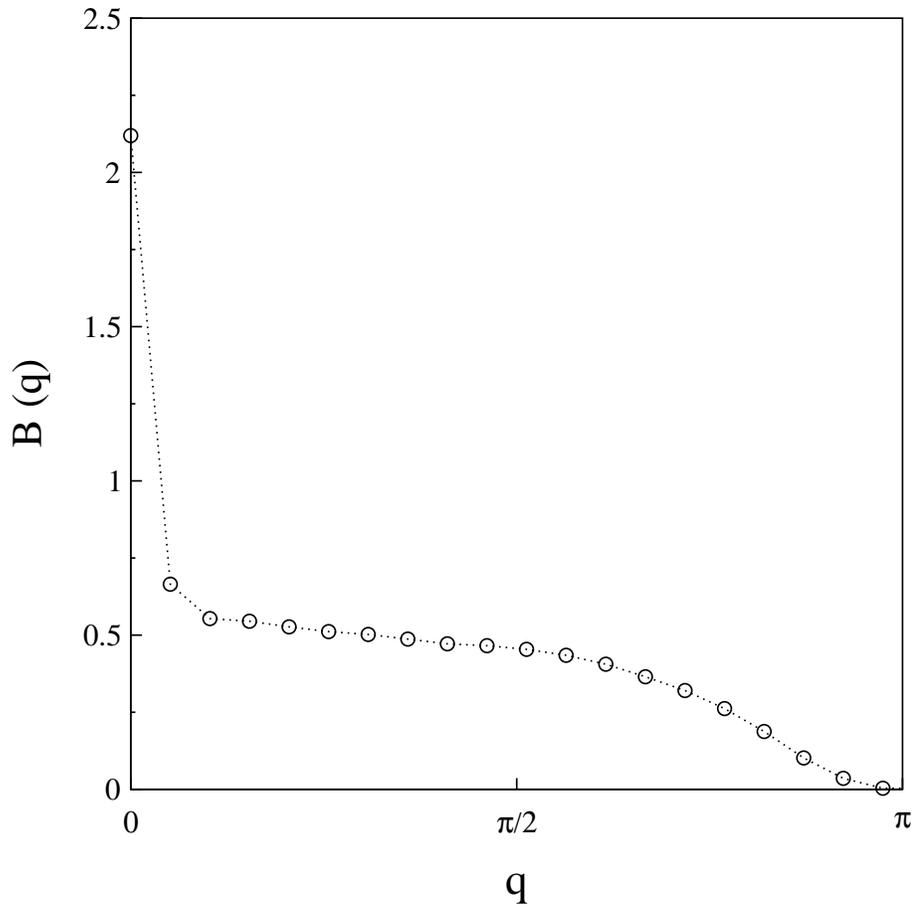}}
\begin{center}
\caption{Bond structure factor corresponding to the bond--bond correlations
 shown in Fig. 7.}
\end{center}
\end{figure}

\clearpage

\begin{figure}
\centerline{\includegraphics[height=4.0cm]{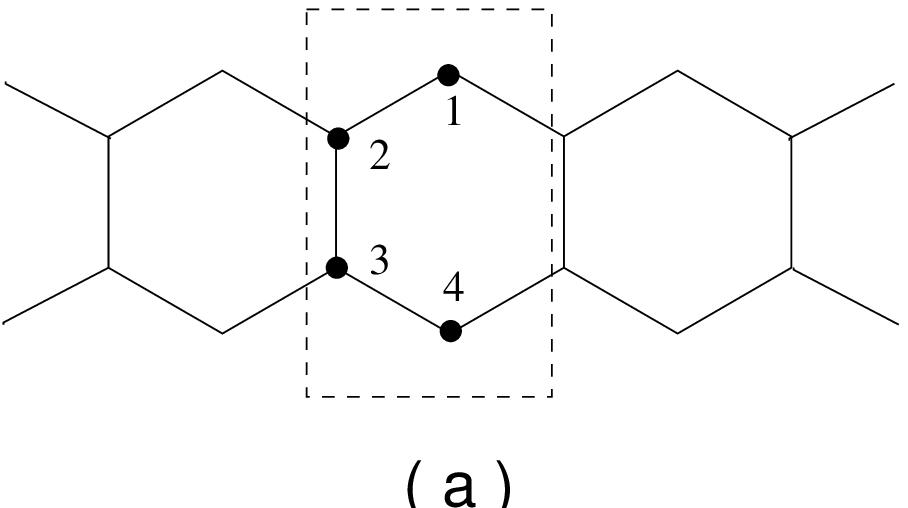}}
\vspace{2.0cm}
\centerline{\includegraphics[height=3.5cm,width=15cm]{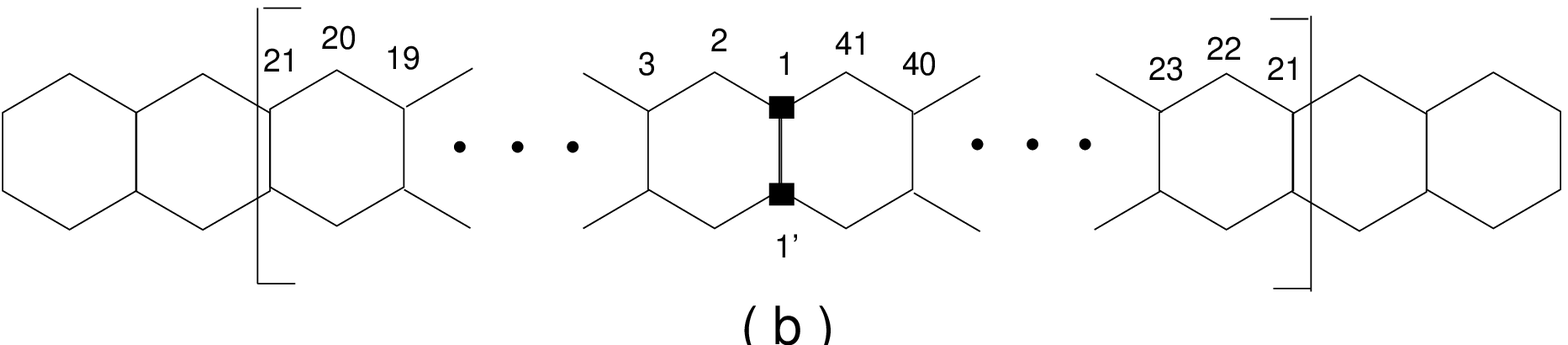}}
\vspace{0.5cm}
\begin{center}
\caption{(a) Unitcell of polyacene showing the different sites, (b) numbering
 of sites in polyacene chain. The new sites are indicated by filled squares. 
 The square brackets indicate the part of the system over which periodic 
 boundary conditions are applied to calculate spin and charge structure 
 factors.}
\end{center}
\end{figure}

\clearpage

\begin{figure}
\centerline{\includegraphics[height=14cm]{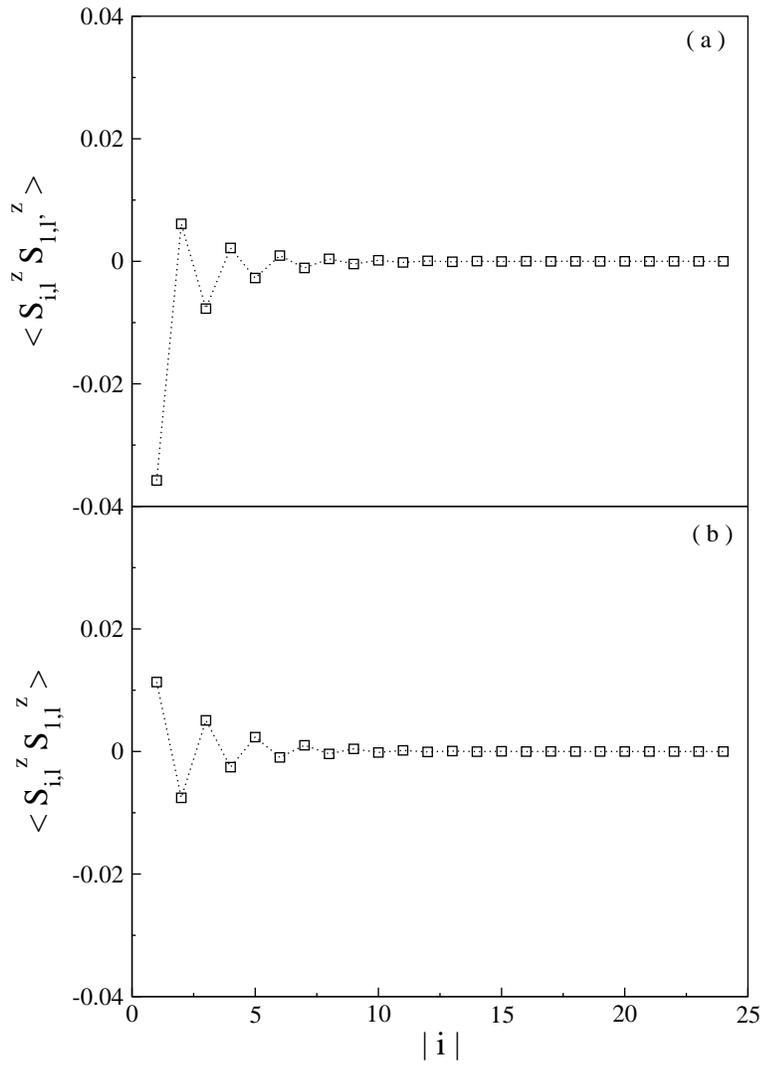}}
\begin{center}
\caption{Spin--spin correlation as a function of the seperation between  the
 new site in the right block and sites on (a) the upper chain and (b) the lower 
 chain of the left block.}
\end{center}
\end{figure}

\begin{figure}
\centerline{\includegraphics[height=14cm]{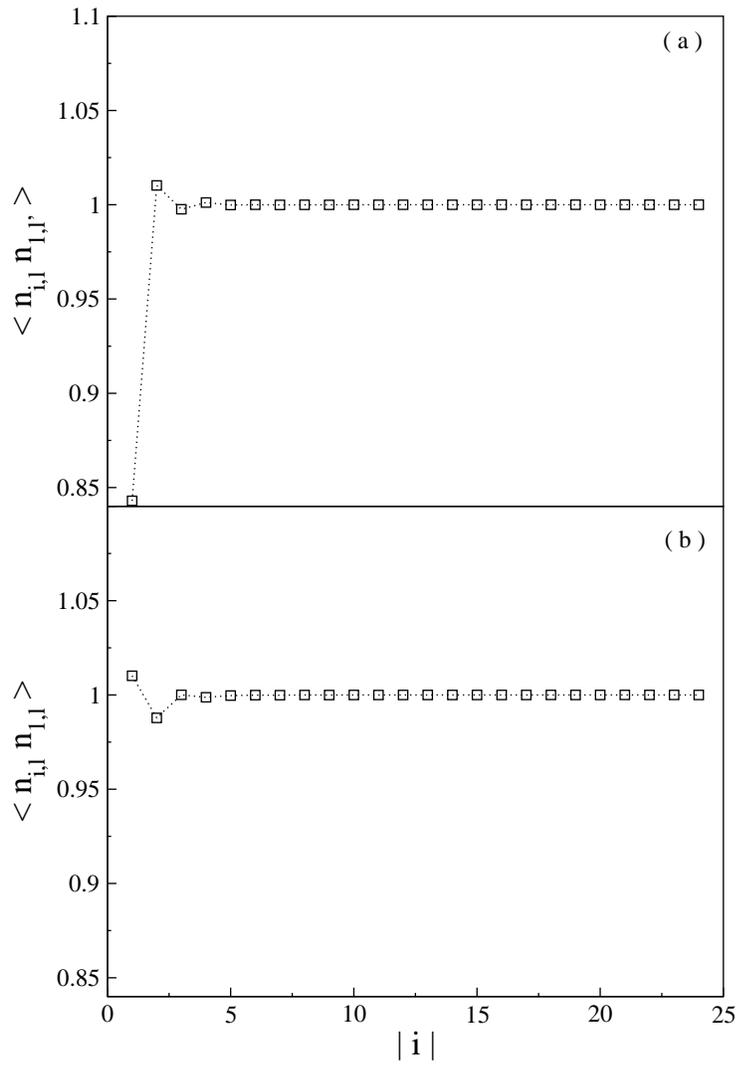}}
\begin{center}
\caption{Charge--charge correlation as a function of the seperation between 
 the new site in the right block and sites on (a) the upper chain and (b) the 
 lower chain of the left block.}
\end{center}
\end{figure}

\begin{figure}
\centerline{\includegraphics[height=12cm]{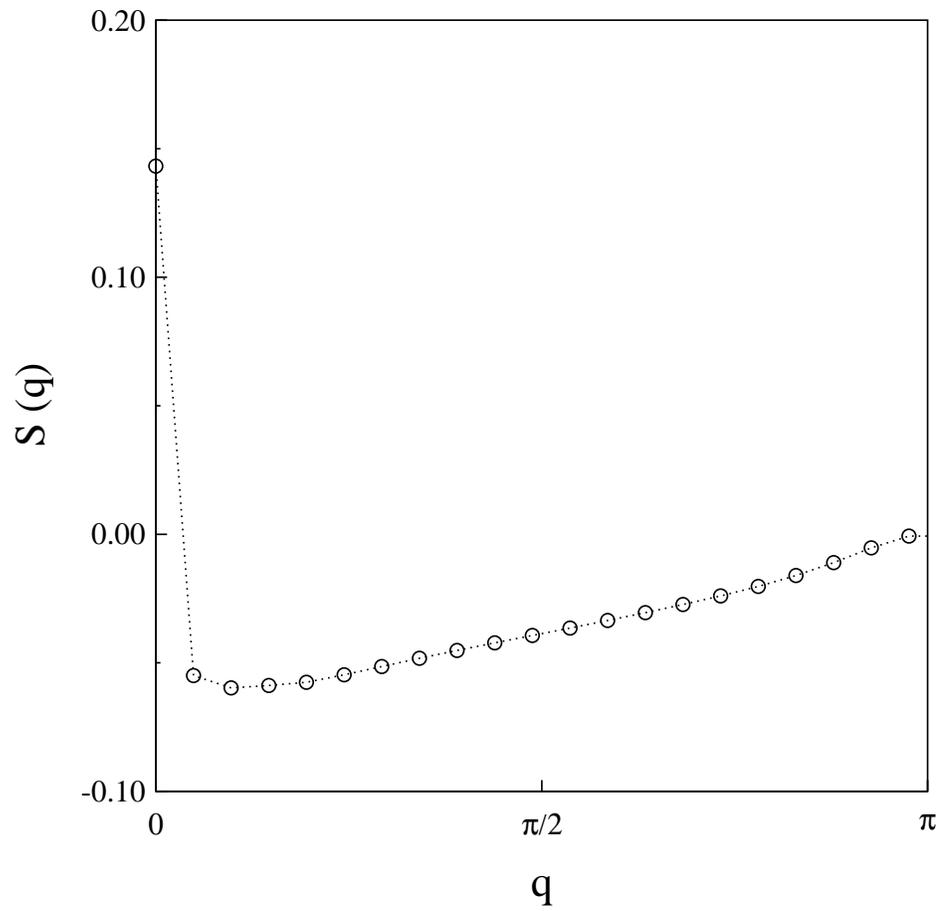}}
\begin{center}
\caption{Spin structure factor obtained from the spin--spin correlation
 function shown in Fig. 10.}
\end{center}
\end{figure}

\begin{figure}
\centerline{\includegraphics[height=12cm]{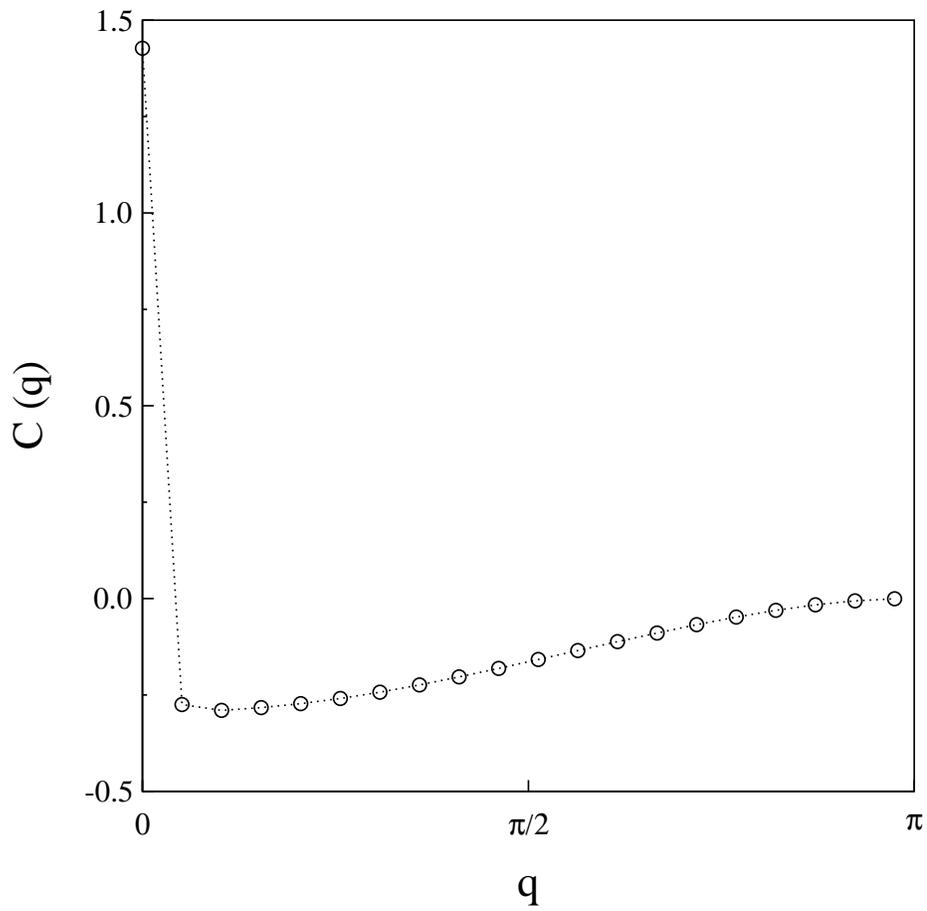}}
\begin{center}
\caption{Charge structure factor obtained from the charge--charge correlation
 function shown in Fig. 11.}
\end{center}
\end{figure}

\end{document}